\documentclass[journal]{IEEEtran}

\usepackage{color}
\usepackage{amsmath}
\usepackage{amsthm}
\usepackage{bm}
\usepackage{amsfonts}
\usepackage{graphicx}
\usepackage{multirow}
\usepackage{url}
\usepackage{hyperref}
\usepackage[flushleft]{threeparttable}

\usepackage{ifpdf} 
\usepackage{cite}
\usepackage{amsmath}
\usepackage{array}
\usepackage[normalem]{ulem}
\usepackage[table]{xcolor}

\usepackage{diagbox}

\begin{document}

\title{Robust Transient Stability Constrained Optimal Power Flow with Power Flow Routers Considering Renewable Uncertainties}

\author{Tianlun~Chen,
        Albert Y.S. Lam,
        Yue~Song,
        and~David J.~Hill
\thanks{T. Chen, A. Y.S. Lam and Y. Song are with the Department of Electrical and Electronic Engineering, The University of Hong Kong, Hong Kong (e-mail: tlchen@eee.hku.hk; ayslam@eee.hku.hk; yuesong@eee.hku.hk).}

\thanks{D. J. Hill is with the Department of Electrical and Electronic Engineering, The University of Hong Kong, Hong Kong, and also with the School of Electrical and Information Engineering, The University of Sydney, Sydney, NSW 2006, Australia (e-mail: dhill@eee.hku.hk; david.hill@sydney.edu.au).}
}

\markboth{}%
{Chen \MakeLowercase{\textit{et al.}}: Transient Stability Constrained Optimal Power Flow with Power Flow Routers and High Penetration of Renewables}


\maketitle

\begin{abstract}
This paper proposes a robust transient stability constrained optimal power flow problem that addresses renewable uncertainties by the coordination of generation re-dispatch and power flow router (PFR) tuning.
PFR refers to a general type of network-side controller that enlarges the feasible region of the OPF problem. The coordination between network-side and generator-side control in the proposed model is more general than the traditional methods which focus on generation dispatch only.
An offline-online solution framework is developed to solve the problem efficiently. Under this framework the original problem is significantly simplified, so that we only need to solve a low-dimensional deterministic problem at the online stage to achieve real-time implementation with a high robustness level. The proposed method is verified on the modified New England 39-bus system. Numerical results demonstrate that the proposed method is efficient and shows good performance on economy and robustness.
\end{abstract}

\begin{IEEEkeywords}
Transient stability, optimal power flow, robust optimization, power flow router, renewable energy.
\end{IEEEkeywords}

\IEEEpeerreviewmaketitle

\section{Introduction} 

\IEEEPARstart{R}{enewable} energy sources (RESs), especially wind power and solar photovoltaic generation, are important components of modern power systems. 
While RESs can bring significant social benefits to the power system, they also introduce enormous technical challenges to power system operators due to their highly variable nature. 
Renewable generation can vary rapidly within very short time periods which requires ancillary techniques and approaches to restore power balance. With the rapid development of electronic-based technology in recent years, FACTS devices such as STATCOMs and Thyristor Controlled Series Capacitors (TCSCs) have been studied for increasing the economic profit and enhancing system security  \cite{haque2004improvement,lu2002static, lin2017optimal}. In \cite{lin2017optimal}, the architecture of power flow router (PFR) was first presented and OPF with PFRs was studied for maximizing system loadability. PFR can be regarded as a new type of series voltage regulator and network side controller which can manage active power and reactive power over transmission lines so that the power system can be more flexible for adapting to uncertainties from renewable generators. Meanwhile, many mathematical methods have been proposed to handle the impact of renewable uncertainties. Stochastic optimization (SO) and robust optimization (RO) approaches, which are typical optimization methods for dealing with uncertainties, have been studied for the power system decision-making process with the consideration of renewable power injections \cite{roald2017chance}.
RO appears to be more reliable since it does not require any distribution model information of RESs and the solution to RO can hedge against any variation of uncertainty parameters. It is natural to apply RO methods for handling renewable uncertainties but generally RO models are quite difficult to solve due to their NP-hardness. To address the problem, many solution methodologies have been studied in the literature \cite{street2010contingency,jiang2011robust,bertsimas2012adaptive,capitanescu2012cautious,
jabr2013adjustable,lorca2017adaptive}. In \cite{bertsimas2012adaptive,street2010contingency,jiang2011robust}, unit commitment is addressed from the view point of robust optimization. In \cite{capitanescu2012cautious}, worst-case analysis is adopted for day-ahead security planning and security-constrained optimal power flow with respect to contingencies.
Reference \cite{jabr2013adjustable} used affine policies where the adjustable generations are affine functions of the renewable power output variations, but it considered DC-OPF models only. In reference \cite{lorca2017adaptive} column-and-constraint generation (CCG) for the robust AC-OPF model is adopted and the paper addressed the nonconvexity via second-order cone programming (SOCP) relaxation. 
The CCG algorithm is an improved version of benders dual cutting plane methods which can solve the two-stage RO problems with lower computational burden. However, it has a strict requirement on explicit expression of constraints. For example, transient stability constraints are designed by trajectory sensitivity analysis in \cite{pizano2014selective} and the sensitivity coefficients cannot be expressed explicitly in terms of variables in the optimization model, which prevents the CCG algorithm from solving the problem. Another way to solve the RO problem is called the scenario approach. The hard constraints with the uncertainty are replaced by a large number of random constraints which correspond to the uncertainty realization. This approach can be generally applied to most robust convex optimization (RCO) problems but the larger number of sampling constraints makes it difficult for online implementation. In this paper, we propose an offline-online solution framework to solve the RO problem which can be used for online implementation.

Another challenge in modern power systems is that the traditional synchronous generators (SGs) have gradually been replaced by renewable generators with less system inertia. These potentially make the power system vulnerable to transient instability.
Transient stability constrained OPF (TSC-OPF) was proposed to dispatch a power system with acceptable cost increase while ensuring its transient stability \cite{gan2000stability,cai2008application,pizano2011new,ledesma2017multi}. It includes algebraic constraints and differential equations, and it is a highly non-convex optimization problem \cite{cai2008application}. Many methods have been designed to tackle this problem and most of them can be categorized into two types: transient energy function (TEF) based and time-domain simulation (TDS) based approaches. The transient stability constraints formulated based on TEF give fast solutions but are often too conservative and currently not suitable for large systems with switching actions \cite{cai2008application}. TDS simulates the dynamics of the system model and the rotor angles can be constrained by TDS-based approach. This approach is advantageous for its accuracy but limited by heavy computational burden especially for large systems with high-order dynamic models. In \cite{pavella2012transient, pizano2011new}, several ways are proposed to improve the computational efficiency while sacrificing accuracy.
Recently, another significant line of work has explored TSC-OPF considering high penetration of RESs.  
Probabilistic approaches \cite{xia2016probabilistic,papadopoulos2016probabilistic} have been recently introduced to transient stability assessment which study the effects of renewable uncertainties. In \cite{en10111926}, a new TSC-OPF model is proposed which includes non-synchronous generation with fault ride-through capability and reactive support during voltage dips. It is revealed in \cite{arredondo2019optimization} that the flywheel energy storage system can be used to improve the system stability while minimizing the generation costs. In \cite{xu2018robust}, a robust dispatch method is proposed to minimize generation fuel cost while maintaining system stability. A small number of testing scenarios were selected to represent the uncertain wind power based on Taguchi's orthogonal array testing.

In this paper, we formulate a robust TSC-OPF problem with PFRs (RTSC-OPF-PFR), minimizing the generation cost while ensuring transient stability under renewable uncertainties. The contributions are reflected in the optimization model and the computation approach. First, we introduce a general type of network-side controller, namely the PFR, into the optimization model which adds further dispatchability from network side for stability enhancement. In contrast to the traditional methods which consider the generation re-dispatch only, the proposed model introduces a coordination of network-side control and generator-side control which has not been explored before. Second, a new offline-online solution framework is designed to enable real time implementation of the proposed problem. With convex relaxation \cite{low2014convex1} and scenario approach \cite{you2018distributed}, the original robust optimization problem is transformed into a convex scenario-based problem with respect to a large number of renewable scenarios. In the offline analysis, scenario reduction method \cite{heitsch2003scenario} and SIME-based transient stability analysis \cite{pavella2012transient} are conducted based on the day-ahead RES prediction interval, which obtain the reduced scenario set and the corresponding linear form of transient stability constraints, respectively. At the online stage, given the short-term prediction interval as a subset of day-ahead prediction interval, we only need to solve a low-dimensional model with respect to a small number of representative renewable scenarios extracted from offline analysis. The offline-online framework significantly reduces the complexity of the original problem.   
The proposed optimization model and framework are verified on the modified New England 39-bus system. The simulation results show the proposed method achieves high robustness level and high computational efficiency at the cost of low economy compared to the traditional TSC-OPF methods when renewable uncertainties are considered. 

\begin{figure}[t] 
\includegraphics[width=\linewidth]{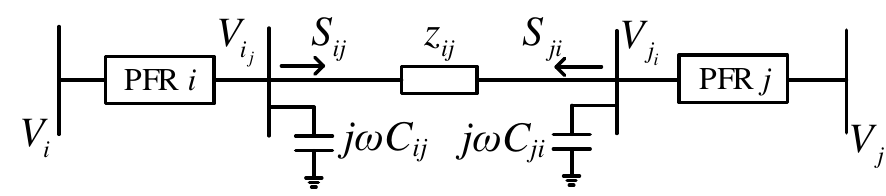}
\caption{Load flow model of PFR on a transmission line \cite{lin2017optimal}. }
\label{fig:PFR_branch}
\end{figure}

The rest of this paper is organized as follows. In Section \ref{sec:problem}, a general RTSC-OPF-PFR problem is presented. The SIME method \cite{pavella2012transient} is used to represent the transient stability constraints in a significantly reduced form. Section \ref{sec:methods} presents the offline-online framework to realize real-time dispatch. Section \ref{sec:simulation} validates the proposed model and framework with case studies. Conclusions and future work are given in Section \ref{sec:conclusion}.

\section{Problem Formulation} \label{sec:problem}
Consider a transmission network represented by an undirected graph $\mathcal{G} := (\mathcal{N}, \mathcal{E})$  with buses $\mathcal{N} := \{1,2,\dots,n\}$ and transmission lines $\mathcal{E} \subseteq \mathcal{N} \times \mathcal{N}$. Then $(i,j)$ and $i \sim j, i \neq j $ are interchangeably used to represent a directed line from $i$ to $j$. Denote $\mathcal{N_G}$ as the set of synchronous generators,  $\mathcal{N_R}$ as the set of buses with renewable farms. Let $Y$ be the power network admittance matrix where both the series and shunt components of transmission lines (see Fig. \ref{fig:PFR_branch}) are included. Then $Y_{ij}$ denotes the $(i,j)$ entry of the admittance matrix $Y$. Let $V_i$ be the complex voltage of bus $i \in \mathcal{N}$. Let $\Omega_j \subseteq \mathcal{N}$ be the neighbors of the bus $i \in \mathcal{N}$ and $E$ be the total number of lines. An example of line $(i,j)$ installed with PFRs $i$ and $j$ is shown in Fig. \ref{fig:PFR_branch}. For PFR $i$,  $V_{i_j} $ denotes the ``branch terminal voltage'' \cite{lin2014architectural} at branch $(i,j)$. For complex quantities $x$, $x^*$ denotes the conjugate transpose.

The renewable generation is modelled by an uncertainty set which includes all the possible scenarios. The expected renewable power output $\hat{P}_{wi}$ at renewable farm $i$ is assumed to be inside a prediction interval $[\underline{P}_{wi},\overline{P}_{wi}]$, which can be obtained based on the historical data or machine learning methods \cite{soman2010review}. We can set $\underline{P}_{wi} = (1-\alpha){P}_{wi}$ and $\overline{P}_{wi} = (1+\alpha){P}_{wi}$, where $\alpha$ is a constant determined by the prediction time horizons and ${P}_{wi}$ is the forecast output at renewable farm $i$ which is also the mean value of the prediction interval. Then denote $\Delta{P}_{wi}$ as the prediction error which belongs to the interval $[-\alpha{P}_{wi}, \alpha{P}_{wi}]$.
All the renewable sources are assumed to operate in the unity power factor mode and can be considered as negative loads.

Let $C_i(P_{G_i})$ = $c_{2i}P_{G_i}^2 + c_{1i}P_{G_i} + c_{0i} $ be the fuel cost of generator $i (i \in \mathcal{N_G})$. $P_{G_i}$ is the active power output of the $i$th generator and $N_G$ is the  number of generators. The objective function can be defined by the expected generation cost as $\sum_{i=1}^{N_G}C_i(P_{G_i})+ \sum_{i=1}^{N_G}c_{2i}'(\rho_i^2)$, where the second term indicates the expected cost function of the recourse policies \cite{jabr2013adjustable} which are employed to balance the generation and load under the realization of uncertainties. Denote $c_{2i}' = [\sum_{j=1}^{N_R}\sum_{k=1}^{N_R}\Lambda_{(j,k)}] \times c_{2i}$ as the coefficient of the generation cost and $\rho_i$ as the participation factor of synchronous generator $i$. The participation factors determine the generation outputs as linear functions of the total change of renewable generation variations with respect to the base case forecast. In this paper, we assign identical participation factors to synchronous generators which balance the RES variations equally. The mean value of the prediction error of renewable power generation is zero and suppose $\Lambda$ denotes the covariance matrix of the prediction errors of renewable farms. We refer to Appendix A in \cite{jabr2013adjustable} for derivation of the expected generation cost and $c_{2i}'$. 

The RTSC-OPF-PFR problem is presented as:
\begin{subequations} \label{robust-TSC-OPF-PFR}
\allowdisplaybreaks[4]
\begin{align} 
& \text{minimize} \quad  \sum_{i=1}^{N_G}C_i(P_{G_i})+ \sum_{i=1}^{N_G}c_{2i}'(\rho_i^2),   \label{TOP-1} \\
& \text{subject to} \quad \forall \hat{P}_{w} \in [\underline{P}_{w},\overline{P}_{w}],  \nonumber\\
\bigg\{ &V_{i_j}=\gamma_{i_j}V_i, \quad \forall (i,j) \in \mathcal{E} \label{TOP-2} \\
&V_{i_j}\sum^n_{j=1}Y_{ij}^*V_{j_i}^* = P_{G_i}+\hat{P}_{wi}-P_{L_i}+{\rm{i}}(Q_{G_i} \nonumber \\
&\qquad\quad\quad -Q_{L_i}), \quad \forall i \in \mathcal{N} \label{TOP-3}\\
&\underline{P}_{Gi}\leq P_{Gi} + \rho_i\sum_{k \in \mathcal{N_R}}\Delta{P}_{wk}  \leq \overline{P}_{Gi}, \quad \forall i \in \mathcal{N_G} \label{TOP-4} \\
&\underline{Q}_{Gi}\leq Q_{Gi}\leq \overline{Q}_{Gi}, \quad \forall i \in\mathcal{N_G}\label{TOP-5} \\
&\underline{V}_{i} \leq |V_{i}| \leq \overline{V}_{i}, \quad \forall i \in \mathcal{N}  \label{TOP-6} \\
&\underline{\gamma}_{i_j} \leq |\gamma_{i_j}| \leq \overline{\gamma}_{i_j},\quad  \forall (i,j)\in \mathcal{E} \label{TOP-7} \\
&\underline{\beta}_{i_j} \leq \angle \gamma_{i_j} \leq \overline{\beta}_{i_j},\quad  \forall (i,j)\in \mathcal{E},  \label{TOP-8}\\
&\sum_{i \in \mathcal{N_G}} \Phi_i(\eta,{P}_{Gi},\hat{P}_{w})(P_{Gi}-{P}_{Gi}^0) + {\eta_0(\hat{P}_{w})} \geq 0, \bigg\} \label{TOP-9}
\end{align}
\end{subequations}
where the objective function \eqref{TOP-1} aims to minimize the expected generation cost.
The RES power forecast interval $\hat{P}_w \in [\underline{P}_{w},\overline{P}_{w}]$ is given by short-term prediction. 
Constraints \eqref{TOP-2} model the branch terminal voltage $V_{i_j}$ of PFR at bus $i$ on branch $(i,j)$. Define $\gamma_{i_j} \in \mathbb{C}$ as the voltage magnitude and phase angle regulation of the PFR $i$. For the lines without PFRs,  $\gamma_{i_j}$ is set as 1.
Constraints \eqref{TOP-3} represent the power balance equations. The branch power flow is controlled by the terminal voltages $V_{i_j}$ and $V_{j_i}$ through the PFRs; the feasible region is larger than for conventional buses without PFRs.


Constraints \eqref{TOP-4}-\eqref{TOP-5} enforce the upper and lower limits of active and reactive power outputs of generators for all the possible renewable power scenarios within the prediction interval.  
Constraints \eqref{TOP-6} and \eqref{TOP-7}-\eqref{TOP-8} refer to the voltage limits and the controllable limits of PFRs.

Constraint \eqref{TOP-9} is the transient stability constraint which is derived from SIME-based analysis \cite{pavella2012transient}. Let $\Phi_i$ be the trajectory sensitivities of each generator $i$. The transient stability margin $\eta$ can be calculated by the difference between the decelerating area and the accelerating area according to the OMIB $P-\delta$ plane in the extended equal-area criterion (EEAC) \cite{xue1989extended}. 
Denote $\eta_0$ as the initial stability margin of the system (i.e., positive for stable condition and negative for unstable condition). Then denote ${P}_{Gi}^0$ and ${P}_{Gi}$ as the active power output at the initial state and after optimization, respectively. Note that ${P}_{Gi}^0$ and ${P}_{Gi}$ are both base case operating points before RES variation compensation.

We further explain the constraint \eqref{TOP-9} as follows.
This constraint refers to the generation re-dispatch under the SIME framework, which has low computational complexity. In general, SIME is an improved version of the EEAC method \cite{xue1989extended}. The system's transient stability is guaranteed by controlling the stability margin of the respective One Machine Infinite Bus (OMIB) equivalent trajectory. Then the constraint can be determined by restricting the critical OMIB trajectory corresponding to the unstable condition \cite{pizano2011new}.
Furthermore, the OMIB's stability margin can be defined as a multi-variable linear function of the mechanical power exchange between generators \cite{pizano2014unified}.
Thus, SIME-based transient stability analysis provides information of OMIB stability margin with defined fault clearing time and mechanical power of the OMIB equivalent. To stabilize a power system which is transiently unstable, generators should be responsible for generation rescheduling at the pre-fault state. The trajectory sensitivities with respect to each generator can be derived as a by-product of TDS analytically or numerically through perturbation analysis \cite{pavella2012transient,xu2016robust}. 
The transient stability constraint \eqref{TOP-9} is determined by realizing the stability margin $\eta \geq 0$ which means the system is transiently stable under a specified contingency.  
For $\forall \hat{P}_{w} \in [\underline{P}_{w},\overline{P}_{w}]$, after compensating the RES variations, the actual generation dispatch at the initial state is ${{P}_{Gi}^{0\prime}} = {P}_{Gi}^0 +\rho_i\sum_{k \in \mathcal{N_R}}\Delta{P}_{wk}$ and the actual generation dispatch after optimization is ${{P}_{Gi}'} = {P}_{Gi} +\rho_i\sum_{k \in \mathcal{N_R}}\Delta{P}_{wk}$. The transient stability constraint can be expressed as 
$\eta_0(\hat{P}_{w}) + \sum_{i \in \mathcal{N_G}}\frac{\partial\eta}{\partial{P}_{Gi}}|_{{{P}_{Gi}^{0\prime}}} ({{P}_{Gi}'}-{{P}_{Gi}^{0\prime}} ) \geq 0$ 
which is simplified as \eqref{TOP-9}. 

In this optimization model, the mechanism of PFRs in enhancing transient stability is mainly reflected by its ability in increasing power transfer capacity. Under the framework of transient stability analysis, the dispatchability of critical generators is restricted since too much power transfer between critical and non-critical generators will reduce the deceleration area. Specifically, generation re-dispatch is to ensure sufficient deceleration area in case of the system contingencies. PFR can increase the available transfer capability of certain critical lines, thus enlarge the feasible region of the OPF problems. In this way, the cost-effective generators can output more power while keeping enough margin for deceleration area, i.e., without violating the transient stability constraints. In general, the network-side control by PFRs further unlocks the generation dispatchability so that the system can achieve the same stability level in a more economic way.


\section{Offline-online Solution Framework} \label{sec:methods}
In this section, we propose the solution methods for the RTSC-OPF-PFR problem. Convex relaxation and scenario approach are adopted to transform the robust optimization model to a scenario-based convex optimization model. A framework consisting of offline analysis and online dispatch is developed to realize the real-time operation.

\subsection{Convex relaxation}
OPF problems are normally non-convex and it is difficult to find a global optimal solution. Inspired by  \cite{low2014convex1}, a convexified RTSC-OPF-PFR problem is formulated.
We introduce $\mathbf{W} := \mathbf{V}\mathbf{V}^* \in \mathbb{C}^{2E \times 2E}$ as the auxiliary matrix, where $\mathbf{V}  \in \mathbb{C}^{2E}$ is defined as the column voltage vector by stacking $V_{i_j},$ and $V_{j_i}$ for branch $(i,j) \in \mathcal{E}$. The diagonal entry of $\mathbf{W}$ is given by $W_{i_ji_j} = |V_{i_j}|^2,\forall i \in \mathcal{N},j\in \Omega_i$ and the off-diagonal entry is given as $W_{i_jk_l} := V_{i_j}V_{k_l}^*, \forall i,j \in \mathcal{N},j\in\Omega_i,l\in\Omega_k$. Denote $W_{i_ji_j} = \Gamma_{i_j} W_i$ where $W_i := |V_i|^2$, $\Gamma_{i_j} := |\gamma_{i_j}|^2, \forall i \in \mathcal{N}, j\in \Omega_i$. The problem \eqref{robust-TSC-OPF-PFR} can be reformulated as follows:
\begin{subequations} \label{relaxed TSC-OPF-PFR}
\allowdisplaybreaks[4]
\begin{align}
&\text{minimize} \quad    \sum_{i=1}^{N_G}C_i(P_{G_i})+ \sum_{i=1}^{N_G}c_{2i}'(\rho_i^2),   \label{relax-1}  \\
& \text{subject to}\quad \forall \hat{P}_{w} \in [\underline{P}_{w},\overline{P}_{w}], \bigg\{
\sum^n_{j=1}Y_{ij}^*W_{i_jj_i} = P_{G_i} \nonumber\\
& +\hat{P}_{wi}-P_{L_i}+ {\rm{i}}(Q_{G_i}-Q_{L_i}),  \quad \forall i \in \mathcal{N} \label{relax-3}\\
& (\underline{V}_{i})^2 \leq W_i \leq (\overline{V}_{i})^2, \quad \forall i \in \mathcal{N} \label{relax-4}\\
&\underline{P}_{Gi}\leq P_{Gi} + \rho_i\sum_{k \in \mathcal{N_R}}\Delta P_{wk}  \leq \overline{P}_{Gi}, \quad \forall i \in \mathcal{N_G} \label{relax-12} \\
&\underline{Q}_{Gi}\leq Q_{Gi}\leq \overline{Q}_{Gi}, \quad \forall i \in\mathcal{N_G}\label{relax-13} \\
& \mathbf{W} \succeq 0,  \label{relax-6}\\
& \text{rank}(\mathbf{W}) = 1,  \label{relax-7}\\
& \underline{\gamma}_{i_j}^2W_i \leq W_{i_ji_j} \leq \overline{\gamma}_{i_j}^2W_i, \forall (i,j)\in \mathcal{E}   \label{relax-8} \\
& \operatorname{Re}\{W_{i_ji_k}\}\tan{\underline{\theta}_{i_ji_k}} \leq \operatorname{Im}\{W_{i_ji_k}\}  \nonumber \\
&\leq \operatorname{Re}\{W_{i_ji_k}\}\tan{\overline{\theta}_{i_ji_k}}, \forall i \in \mathcal{N}, j\neq k \in \Omega_i, \label{relax-9}\\
&\operatorname{Re}\{W_{i_ji_k}\} \geq \underline{\gamma}_{i_j}\underline{\gamma}_{i_k}W_i\cos{(\max\{|\underline{\theta}_{i_ji_k}|,|\overline{\theta}_{i_ji_k}|\})},  \nonumber \\
&\qquad\qquad\qquad\qquad\qquad\qquad \forall i \in \mathcal{N}, j\neq k \in \Omega_i, \label{relax-10}\\
&\sum_{i \in \mathcal{N_G}} \Phi_i(\eta,{P}_{Gi},\hat{P}_{w})(P_{Gi}-{P}_{Gi}^0) + \eta_0(\hat{P}_{w}) \geq 0, \bigg\} \label{relax-11}
\end{align}
\end{subequations}
where \eqref{relax-3} gives the power flow constraints and \eqref{relax-4} gives the voltage limits. Constraints \eqref{relax-8}-\eqref{relax-10} are convex which account for the equal feasible range of non-convex constraints \eqref{TOP-2}-\eqref{TOP-3},\eqref{TOP-7}. Denote $\underline{\theta}_{i_ji_k}$ and $\overline{\theta}_{i_ji_k}$ as the lower and upper limits of angular difference between two terminal voltages $V_{i_j}$ and $V_{i_k}$.
Constraint \eqref{relax-6} determines that the matrix $\mathbf{W}$ is positive semi-definite.
Removing the rank constraint \eqref{relax-7} makes problem \eqref{relaxed TSC-OPF-PFR} convex. Numerical
case studies indicate that SDP relaxation tends to be
exact in many cases \cite{low2014convex1} and the optimal solution of original nonconvex problem can be recovered from the optimal solution of its relaxed form. A penalty which can be the total reactive power generation or the apparent power loss can be added in the objective function to guarantee the rank-one solution while having negligible impact on the objective value. A tree decomposition method can be adopted to reduce the computational burden of SDP relaxation in large-scale power systems \cite{lin2017optimal}.
By using the SDP relaxation, we can adopt the scenario approach \cite{you2018distributed} to solve problem \eqref{robust-TSC-OPF-PFR}.

\subsection{Scenario Approach} \label{subsec: SP}
The RCO problem \eqref{relaxed TSC-OPF-PFR} can be written in a compact notation as follows:
\begin{equation} \label{eq:RCO}
    \min_{\theta \in \Theta} \; C(\theta) \quad \text{subject to} \; f(\theta,\xi) \leq 0, \quad \forall \xi \in \Xi,   
\end{equation}
where $\Theta \subseteq \mathbb{R}^n $ is a closed set with decision variables and $C$ is the cost function. Constraints $f(\theta,\xi) $ are convex in the decision vector $\theta$ for $\forall \xi \in \Xi$, where $\xi$ is the system uncertainty (in this paper the renewable power $\hat{P}_w$) according to an arbitrary absolutely continuous distribution $\mathbb{P(\cdot)}$ over $\Xi$. RCO is NP-hard if $f(\theta,\xi) $ is nonlinear and it is difficult to obtain the worst case solution which is still conservative in most cases. A scenario approach is proposed to transform the RCO into a scenario problem \cite{you2018distributed}. The basic idea is to replace the hard constraint in \eqref{eq:RCO} by a probabilistic approximation.
Let $\xi^1,\xi^2, ..., \xi^n$ be the samples which are independently generated from the distribution $\mathbb{P_{\xi}}$. The constraint can be replaced by a set of random inequality constraints as below:
\begin{equation} \label{eq:SP constraints}
\bigcap_{i=1}^{N}\{\xi|f(\theta,\xi^{(i)}) \leq 0 \},
\end{equation}
where $N$ is the number of the constraints. Then the scenario problem is formulated under the constraints \eqref{eq:SP constraints}:
\begin{equation} \label{eq:SP}
    \min_{\theta \in \Theta} \; C(\theta) \quad \text{subject to} \; f(\theta,\xi^{(i)}) \leq 0  \quad i = 1, ..., N \in \Xi   
\end{equation}
In this way, \eqref{eq:RCO} can be approximately addressed by solving \eqref{eq:SP} if the sample complexity $N$ satisfies the following condition:
\begin{equation} \label{eq:SP lemma}
   N \geq \frac{e}{\epsilon(e-1)}(-\ln{\delta}+n-1),
\end{equation}
which is verified by \textit{Lemma 1} in \cite{you2018distributed}.

\begin{figure}[!t]  
\includegraphics[width=\linewidth]{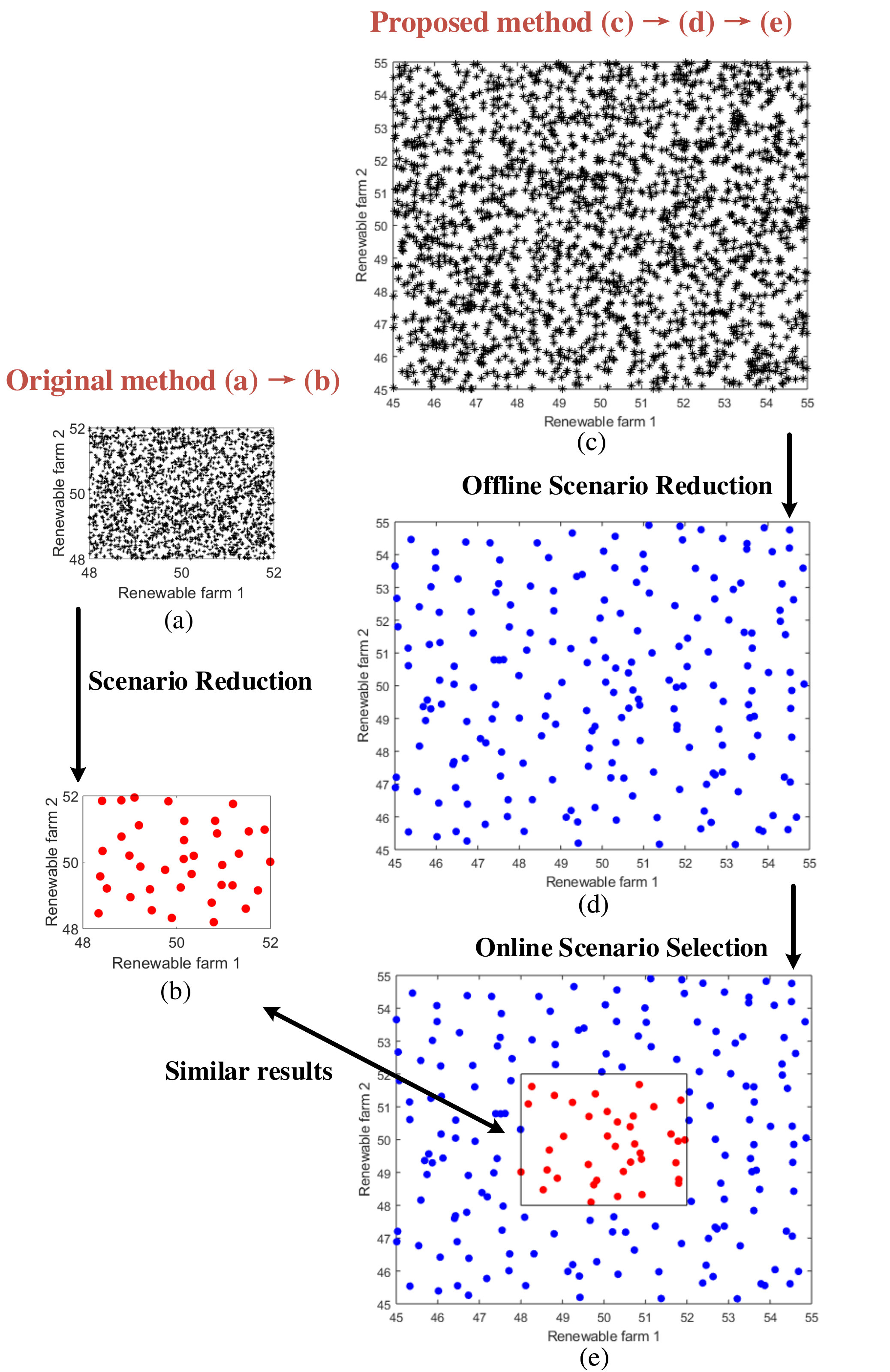}
\caption{Offline-online scenario reduction and selection}
\label{fig:Offline_reduction}
\end{figure}

For the uncertainty introduced by the renewable energy generators in  RTSC-OPF-PFR, the above scenario approach is adopted and RCO is computationally tractable by the sampling approximation.
To address the RCO problem \eqref{relaxed TSC-OPF-PFR} via the scenario approach, a large number of renewable scenarios are sampled following the uniform distribution over the short-term (e.g., 10 mins) forecast interval $\Xi_{st}:= \{\hat{P}_w \in \mathbf{R^r}|\underline{P}^{st}_{wi} \leq \hat{P}_{wi} \leq  \overline{P}^{st}_{wi}\}$. An example is shown in Fig. \ref{fig:Offline_reduction}(a) where the 5000 scenarios are generated based on the forecast interval. For such a large sampling set, the computational burden of the problem becomes too high and cannot be solved efficiently. Scenario reduction methods are used to reduce the number of scenarios while keeping the stochastic information as complete as possible \cite{heitsch2003scenario}. The reduced scenarios are shown in Fig.  
\ref{fig:Offline_reduction}(b). However, although we can solve the scenario problem in \eqref{eq:SP} with the reduced scenarios, there remains at least two points that prevent the proposed model from online implementation: (a) scenario reduction will take a long time if the original sampling set is large; (b) transient stability constraints are derived from TDS and every scenario corresponds to one TDS. Even with a reduced number of scenarios the computational complexity of TDS is very high. In short, online dispatch is not directly applicable as the samples extracted from the short-term forecast interval are not available and the transient stability analysis cannot be guaranteed to be done within several minutes especially for large systems with high-order models of system components. 

As an alternative, we propose an offline-online framework to include the scenario reduction and transient stability analysis in the offline process. Then an online dispatch can be achieved by directly selecting a minority of the most representative scenarios from the offline database. In this way, the transient stability constraints used in the online dispatch can be approximated by the offline analysis and the RO problem can be solved efficiently for online implementation.

\subsection{Assumptions for Offline-online Solution Framework}  \label{subsec: Assumption}
The assumptions adopted for the offline-online solution framework to be established are listed below: \\
(A1) The scenario reduction method can be applied to the renewable power scenarios which are used to approximate the transient stability constraints within the renewable uncertainty interval given by short-term prediction, which results in low errors in the robustness level of the solutions.
\\
(A2) The RES power uncertainty interval given by short-term prediction is a subset of that given by day-ahead prediction;

In the following we further discuss the reasonableness of these assumptions:

For (A1), renewable power scenarios based on the short-term prediction interval are used for transient stability analysis and generating the transient stability constraints. 
The sufficiently large number of renewable power scenarios required by the scenario approach can be reduced into a much smaller number of scenarios while still preserving the random data information in it as complete as possible \cite{heitsch2003scenario}. This reduction is also validated by the fact that many renewable power scenarios are physically similar and the corresponding trajectory sensitivities (which forms the transient stability constraints) are very close. Thus, the scenario reduction approach can be applied to simplify the sample complexity by selecting a minority of the most representative scenarios from the original set. This results in lower computational complexity and negligible errors in the optimization solutions. This is acceptable for practical engineering applications such as TSC-OPF.

For (A2), this assumption is trivial as the short-term prediction is expected to be more accurate than the day-ahead one. The introduction of day-ahead prediction is due to the following reasons.
Since the scenario reduction process and transient stability analysis (even for a small number of scenarios) are both time consuming, an efficient offline-online framework is developed to enable the method for real time implementation. This framework makes use of both short-term prediction and day-ahead prediction, which will be detailed in Section \ref{subsec: offline} and Section \ref{subsec: online}.

Note that other methods such as CCG algorithm are also able to efficiently solve the RO problem. However, we adopt the SIME-based transient stability analysis in this paper. Transient stability constraints are designed based on the trajectory sensitivity coefficients of the stability margin with respect to mechanical power of generators. The relationship between trajectory sensitivity coefficients and other variables such as renewable uncertainties cannot be expressed in an explicit form, which renders general RO methods difficult for solving the proposed model with transient stability constraints. 
The scenario approach is more suitable for online implementation and the following sections demonstrate how to achieve it.

\subsection{Offline Analysis} \label{subsec: offline}
In what follows, we firstly present an offline analysis which includes scenario reduction and transient stability assessment for the reduced scenarios. 

The offline analysis starts from the day-ahead forecast of renewables. As stated in Section \ref{sec:problem}, the variable renewable power output is modelled by an uncertainty set $\Xi$ with a prediction interval $\Xi_{da} := \{\hat{P}_w \in \mathbf{R^r} |  \underline{P}^{d}_{wi} \leq \hat{P}_{wi} \leq \overline{P}^{d}_{wi} \}$.
Since the forecast values of RESs can be inaccurate, a large number of renewable power scenarios are generated according to the prediction interval, which means any value within the prediction interval can be the realization of RESs. 
As shown above, we use a scenario approach instead of obtaining the worst-case solutions within the uncertainty interval. A set of random constraints are generated to approximate the hard constraint in problem \eqref{eq:RCO}. 
Each uncertainty sample should be extracted from an absolutely continuous distribution over the uncertainty set. In this paper, each renewable power output is independently extracted from a uniform distribution based on Monte Carlo simulation (MCS).
With such a large number of random scenarios, the computational burden will still dramatically increase. In addition, the solutions will be too conservative if we directly solve the scenario problem based on the day-ahead forecast interval. It is necessary to reduce the initial scenarios to a smaller set.
In other words, using a proper scenario reduction technique will downsize the scenario set while deriving the optimal solution close to that obtained from the original scenario set. Many algorithms have been proposed and the fast forward reduction method is one of the most popular and accurate reduction algorithms \cite{heitsch2003scenario}. We denote $W$ as the initial scenario measure and $Z$ as the preserved scenario measure which is closest to $W$ in terms of probability distances.
The Kantorovich distance $D_K$ between of $W$ and $Z$ is calculated as stated in \cite{heitsch2003scenario}.
The optimal selection of a single scenario will be repeated until a predefined number of scenarios are preserved and the original scenario measure $W$ is reduced to measure $Z$. 
The offline scenario generation and reduction is shown in Fig. \ref{fig:Offline_reduction} (c) and (d).
By now we have obtained a reduced scenario set based on day-ahead forecast interval and the number of scenarios is acceptable for transient stability analysis in the offline process.
Then we conduct TDS with the contingencies and the renewable power generation from the reduced scenario set. If the system is transiently unstable for the initial operating point, a transient stability constraint is constructed based on trajectory sensitivity analysis and the sensitivity coefficients of generators with respect to each scenario are obtained. After the above steps, an offline database is established and will be used for online dispatch.

\subsection{Online Dispatch} \label{subsec: online}
\begin{figure}[!t] 
 \includegraphics[width=3.4in]{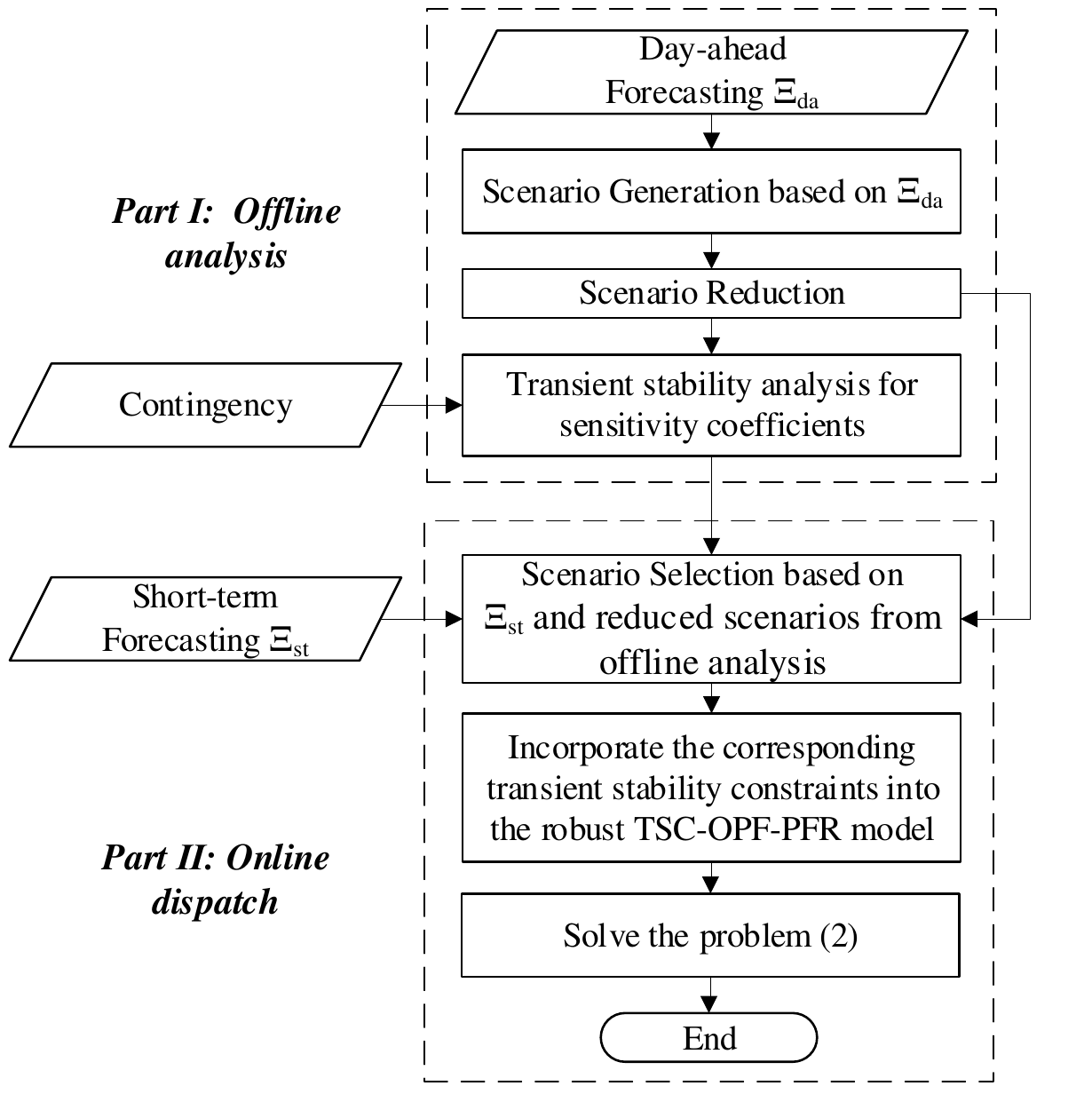} 
 \caption{Offline-online solution framework flowchart}
 \label{fig:flowchart}
\end{figure}

With the day-ahead database constructed as above, the online dispatch can be applied as follows.
In each day-ahead cycle, a short-term (e.g., 10 mins) renewable power forecasting is used to determine the online selection interval in the next instant. With the state-of-the-art technology  in renewable power prediction \cite{soman2010review}, in general, the prediction errors increase at higher time horizons because of the decrease in accuracy \cite{pappala2009stochastic}. Under the assumption (A2), the short-term forecast interval is within the range of day-ahead interval: $\Xi_{st}:= \{\hat{P}_w \in \mathbf{R^r} | \underline{P}^{st}_{wi} \leq \hat{P}_{wi} \leq \overline{P}^{st}_{wi}, \underline{P}^{st}_{wi} > \underline{P}^{d}_{wi}, \overline{P}^{st}_{wi} < \overline{P}^{d}_{wi}  \}$.
Based on the short-term forecast interval ${\Xi}_{st}$, we can select the representative scenarios, as in Fig. \ref{fig:Offline_reduction}(e), from the reduced scenario set, as in Fig. \ref{fig:Offline_reduction}(d). Since the scenarios are all independently extracted via a uniform distribution over the uncertainty set, we can regard the original set in Fig. \ref{fig:Offline_reduction}(a) as a part of original set in Fig. \ref{fig:Offline_reduction}(c). Thus the reduced scenario set Fig. \ref{fig:Offline_reduction}(b) can also be regarded as giving similar results to the scenario set inside the red box in Fig.  \ref{fig:Offline_reduction}(e), which means only the scenarios within the short-term forecast interval $\Xi_{st}$ are selected. This equivalence is based on the same original scenario set and its effectiveness will be verified in the case studies.
So far we can construct the most representative scenario set by the offline-online framework which is less conservative than directly using the day-ahead prediction interval for the robust optimization problem. This scenario set is similar to the one which is directly derived from the short-term prediction interval and it does not require conducting scenario reduction or transient stability analysis in the online dispatch. 
In this way, the problem in Section \ref{subsec: SP} can be addressed for online implementation. Only a small number of scenarios are finally selected for online dispatch and the scenario reduction is finished offline. The transient stability constraints are also constructed based on the trajectory sensitivities in the offline database so that no TDS is done during the online dispatch. The main procedures for solving the RTSC-OPF-PFR problem are depicted in Fig. \ref{fig:flowchart}.

\textit{Remark 1:} With the development of convex relaxation techniques \cite{low2014convex1}, matrix decomposition techniques \cite{molzahn2013implementation}, and optimization modelling tools like CVX \cite{cvx}, practically sized systems are computationally tractable for OPF problems. In online dispatch, the computational complexity depends on solving the RTSC-OPF-PFR problem \eqref{relaxed TSC-OPF-PFR} (which excludes the transient stability analysis). The problem \eqref{relaxed TSC-OPF-PFR} is fundamentally a static OPF problem in convexified form with extra linear transient stability constraints. For static OPF problems, the above computational advances can be adopted to accelerate the computation which makes it possible for online dispatch even for a large-scale system with hundreds of power plants, e.g., OPF on Polish system can be accomplished in the order of minutes \cite{molzahn2013implementation}. Thus, it is reasonable to affirm that the proposed methodology can be applied for online implementation in large-scale practical systems.

\section{Numerical Results} \label{sec:simulation}
To evaluate the proposed model and the solution framework, a modified New England 39-bus system with wind farms (WFs) is firstly prepared in the numerical examples. Three synchronous generators G32, G37, G39 are replaced by wind power generators. 
PFRs are installed at lines 16-21, 16-24, 26-27 which are close to the fault locations or between the generators and fault locations.\footnote{The optimal allocation of PFRs is beyond the scope of this paper and will be considered in the future.} 
Day-ahead and short-term forecasts are used for uncertainty modelling. The wind power generation is assumed to follow the uniform distribution with no correlation between both wind farms and all the possible scenarios within the forecast interval are considered. Each scenario is treated equally in the offline analysis including the worst case and this assumption also satisfies the requirements of robust design in Section \ref{subsec: Assumption}.
Then $\alpha$ is set as $0.1$ for day-ahead forecasting and $0.03$ for short-term forecasting, which shows the latter is more accurate than the former. For simplicity, we ensure the short-term interval is within the day-ahead interval and the mean values of two intervals are the same.
We assume the mean values of the wind power output are equal to the original generation output of the replaced synchronous generators.    

For synchronous generators, the subtransient fourth-order generator model and the IEEE Type-II exciter are used \cite{milano2005open}. The capacities of the original synchronous generators are increased by 1.2 times for accommodating the wind power uncertainties. Doubly fed induction generators (DFIGs) are adopted in this paper for modelling the wind farm dynamics \cite{milano2005open}. The rotor angle trajectories of DFIGs are not used in the OMIB construction and their effects on the transient stability are reflected on the trajectories of the synchronous generators. For the load type, all the loads are considered as frequency dependent loads \cite{milano2005open} in the time-domain simulation for the offline SIME-based analysis and the online dispatch verification. Since the frequency deviation is always zero in the steady state, this type of load reduces to constant power load which is consistent with our OPF model.

Voltage magnitude limits for all nodes are set to $[0.95, 1.05]$ p.u. 
The system data and dynamical data are adopted from \cite{basecase39,zimmerman2011matpower}.
The parameter specifications follow  \cite{lin2017optimal}, particularly  $\underline{\gamma}_{i_j} = 0.95, \overline{\gamma}_{i_j} = 1.05, \overline{\beta}_{i_j} =- \underline{\beta}_{i_j} = 10^{\text{o}} $. For the scenario problem, we set $\epsilon = 0.005$ and $\delta = 0.001$ and determine $N \geq 4084$ from \eqref{eq:SP lemma}. 
The simulation is conducted on a 64-bit PC with 3.2 GHz CPU and 16 GB RAM. TDS and SIME-based analysis are performed in the MATLAB platform. The optimization problem is solved by SDPT3 via CVX toolbox \cite{grant2014cvx}.

\begin{figure}[t] 
 \includegraphics[width=3.5in]{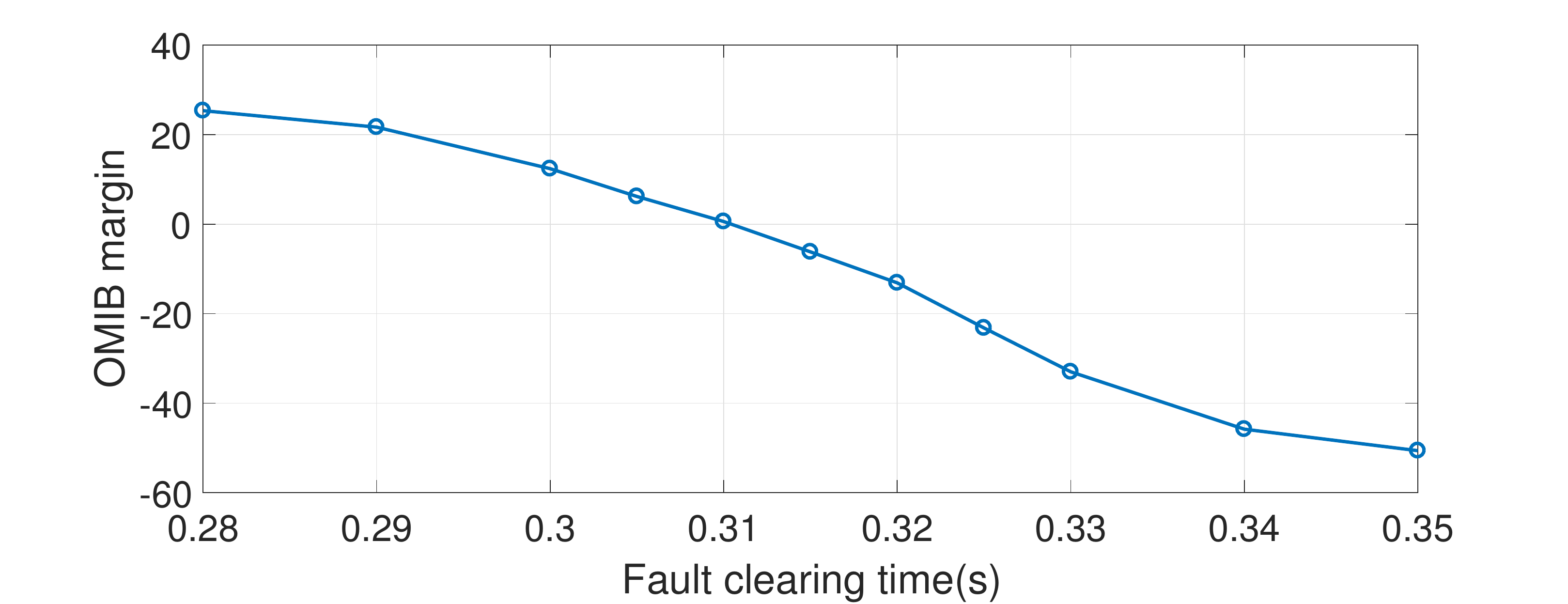}
 \caption{Margin v.s. Fault Clearing Time}
 \label{fig:CCTvsMargin}
\end{figure}

\begin{table}[t] 
\renewcommand{\arraystretch}{1.3} 
\begin{threeparttable}
\centering
\caption{Generation dispatch for New England System (MW)} 
\label{table:1}
 \begin{tabular}{ c| c| c| c| c| c }
\hline \hline
 Approach  & 30 & 31 & 32* & 33 & 34    \\ 
  \hline
Normal OPF($P_G^0$)  & 591.91& 595.12 & 650.00& 580.68& 580.19    \\   
\hline
RTSC-OPF-PFR($P_G$)  & 629.37  & 635.37 & 650.00 & 614.29 & 609.38  \\
\hline \hline
 Approach & 35 & 36 & 37* & 38 & 39*   \\ 
  \hline
Normal OPF($P_G^0$)  & 585.82& 582.55& 540.00& 583.19& 1000.00    \\   
\hline
RTSC-OPF-PFR($P_G$)  & 616.98& 616.52& 540.00 & 379.20& 1000.00  \\
\hline \hline
Total cost change &\multicolumn{2}{|c|}{+2.060\%} & Robustness & \multicolumn{2}{|c}{100\%}   \\ 
\hline \hline
\end{tabular}
\begin{tablenotes}
      \footnotesize
      \item *Wind farms (non-controllable)
    \end{tablenotes}
\end{threeparttable}
\end{table}

\subsection{Base case for New England System} 
\begin{figure}[t]   
  \includegraphics[width=3.5in]{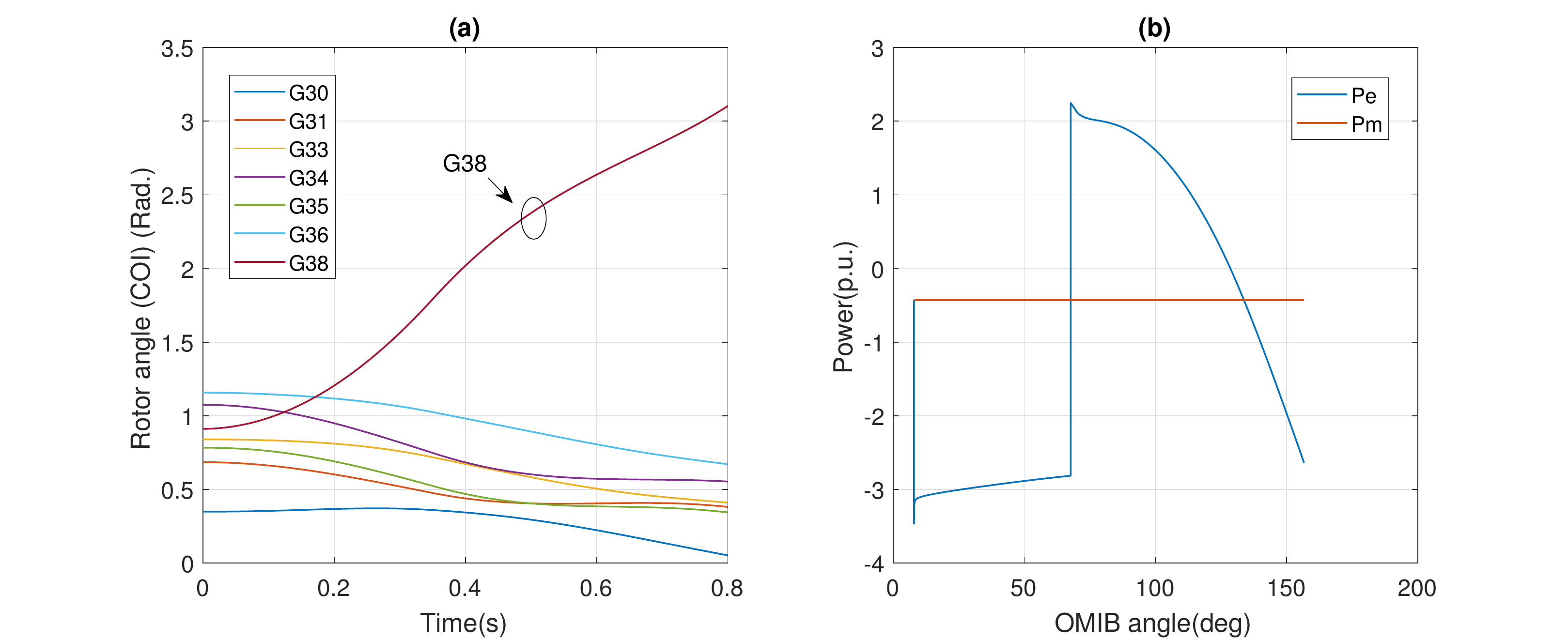}
 \includegraphics[width=3.5in]{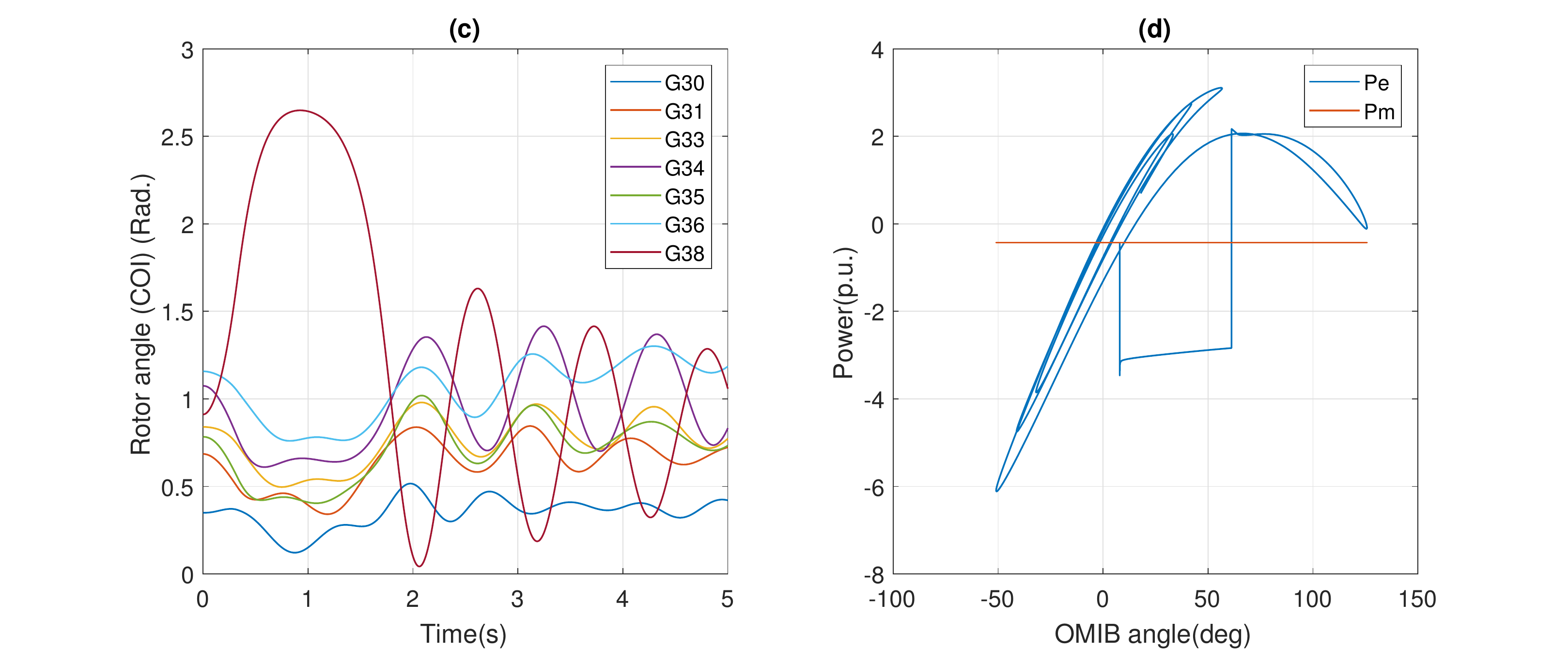}
 \caption{Rotor angle trajectories and Pe-OMIB angle plane under base case: (a)-(b) unstable under initial clearing time; (c)-(d) stable under CCT}
 \label{fig:Base case}
\end{figure}

Before considering the PFRs, wind power uncertainties and transient stability constraints, an initial optimal operating point is calculated from the base case by solving original OPF and the results are shown in Table \ref{table:1}. A single contingency is defined by a solid three-phase fault which begins at $t_0 = $ 0 ms at bus 29 and cleared at  $t_{cl} =$ 350 ms by tripping the line 26-29. We conduct a TDS using the base case dispatch as the initial operating point. As shown in Fig. \ref{fig:Base case} (a) and (b), G38 loses its synchronism with respect to the rest of the system and $P_e$ crosses $P_m$ which indicates the system is transiently unstable.
The critical clearing time (CCT) can be obtained from extra TDS based on the quasi-linear  relationship of stability margin with respect to fault clearing time, shown in Fig. \ref{fig:CCTvsMargin}. In this figure, CCT can be estimated as 310 ms. Under the CCT, the rotor angle trajectories and $P_e$-OMIB plane are given in Fig. \ref{fig:Base case} (c) and (d). As Fig. \ref{fig:Base case} (d) shows, $P_e$ curve returns back before crossing $P_m$, which indicate that the system is transiently stable under the CCT for the above contingency.
Note that this estimation method is adequate for practical use because the linearity around the zero OMIB stability margin is very strong.   
We test the system transient stability by incorporating variable wind power scenarios. A robustness degree in \cite{xu2018robust} is adopted to evaluate the robustness of the base case dispatch with 1,000 scenarios randomly generated from the forecast interval. With the given contingency, the system is transiently unstable for 90.7\% of the scenarios. So it is necessary to derive a more robust framework for dispatching a high-level wind power penetrated system.

\subsection{Single Contingency Case for New England System}
We consider the same contingency as stated in the base case. To verify the proposed solution framework, the whole process for maintaining the system stability under wind power uncertainties is discussed as follows. Based on the offline day-ahead prediction, the wind power generation at each WF can be forecasted in the next day. We can obtain the mean value and the interval of the uniform distribution where $\alpha$ is set as 10\%. And 5000 wind power scenarios are randomly generated following the uniform distribution function. Then fast forward reduction method is used for scenario reduction from 5000 to 200. The 200 wind power scenarios are all used to construct the offline database. For each wind scenario within the reduced set, TDS is carried out with the base operating point. SIME-based analysis and trajectory sensitivity analysis are performed to derive the transient stability constraint if the system is transiently unstable under the initial operating point. 
According to the proposed methodology, the main purpose of this computation process is to build an offline database for realizing online dispatch. 
Then, in the online analysis process, a short-term forecasting is performed for more accurate forecast values of wind power generation where $\alpha$ is set as 3\%. With the solution methodology discussed in Section \ref{sec:methods}, only 54 scenarios are finally selected based on the short-term forecast interval. The corresponding 54 scenarios are used for constructing the scenario problem and finally we solve the scenario problem to approximate the solutions of RTSC-OPF-PFR. These determine the optimal re-dispatch to maintain system transient stability after the contingency. The adjustment with respect to the base case dispatch is shown in Table \ref{table:1}. It can be observed that the generation of G38 decreases while those of the remaining generators increase compared to the base operating point, owing to the positive sensitivities of G38 and negative sensitivities of the remaining generators. The robustness of the proposed robust TSC-OPF-PFR is 100\% which verifies the effectiveness of the proposed methodology for providing a more robust transiently stable operating point.
Table \ref{table:1} also shows the proposed approach only introduces 2.060\% increase of total generation cost with respect to the base case, which means the robustness can be achieved with a low additional cost. The total computation time for online dispatch is 49.81 s, indicating the proposed scheme allows real-time implementation.

\begin{table*}[htbp!] 
\renewcommand{\arraystretch}{1.3} 
\begin{threeparttable}
\centering
\caption{Numerical results under multi-contingencies}
\label{table:2}
 \begin{tabular}{  m{1in} m{1in} m{1in} m{1in} m{1in} m{1in}  }
 \hline \hline
 & Normal OPF  & TSC-OPF \cite{pizano2011new} & TSC-OPF-PFR \cite{chen2018transient} & RTSC-OPF &  RTSC-OPF-PFR \\ 
  \hline
Cost (\$/hr) &25241.21  & 27267.39 & 26899.03  & 28431.44	&27580.16   \\ 
  \hline
Average CPU time (s) &20 &  20	& 50  & 20	& 50     \\  \hline
Robustness under C1 &9.3\%  & 41.6\% & 43.5\%      &99.4\%	    &100\%    \\
  \hline
Robustness under C2 &10.1\%  & 22.0\% & 35.5\%      &98.6\%	    &99.9\%    \\
  \hline
Robustness under C3 &15.6\%  & 58.5\% & 67.9\%      &97.2\%	    &99.7\%    \\
 \hline
CCTs* under C1 (s) &0.310 & 0.615 & 0.653      &0.701    &0.724    \\
  \hline
CCTs* under C2 (s) &0.313  & 0.675 & 0.785      &0.843	    &0.889    \\
  \hline
CCTs* under C3 (s) &0.457  & 0.688 &  0.743      & 0.836	    &0.928    \\
\hline \hline
\end{tabular}
\begin{tablenotes}
      \footnotesize
      \item *All the CCTs are obtained under the forecast wind power scenario.
    \end{tablenotes}
\end{threeparttable}
\end{table*}

\subsection{Multi-Contingency Case for New England System}

\begin{figure}[t] 
 \includegraphics[width=\linewidth]{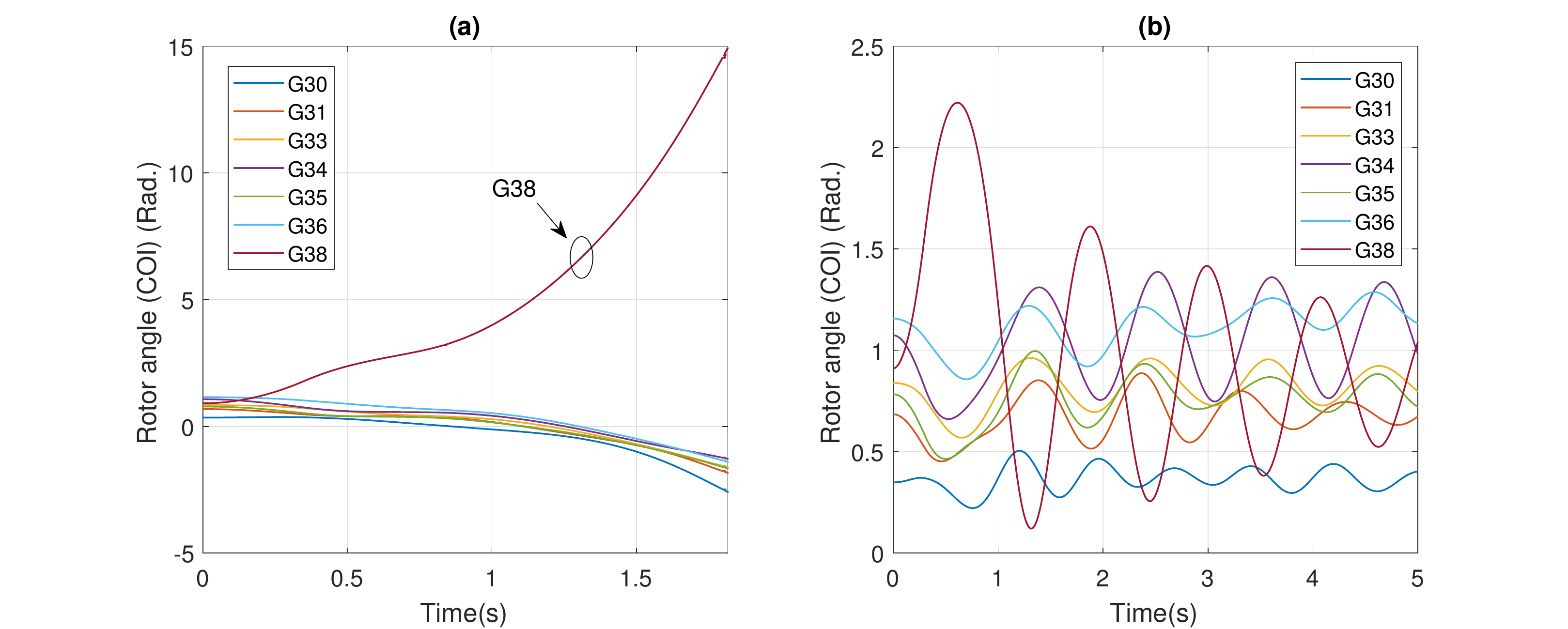} 
 \caption{Rotor angle trajectories under Contingency A: (a) unstable under original OPF; (b) stable under RTSC-OPF-PFR}
 \label{fig:L26-29 stable}
\end{figure}

\begin{figure}[t] 
 \includegraphics[width=\linewidth]{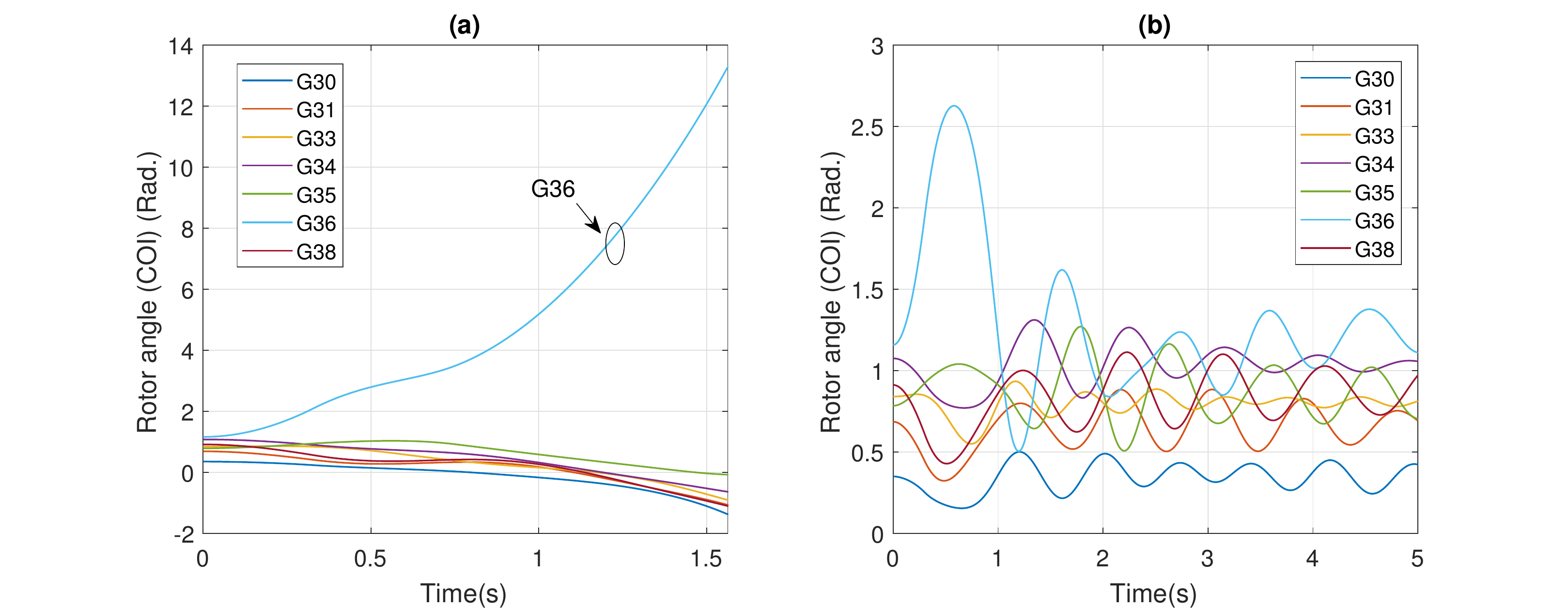} 
 \caption{Rotor angle trajectories under Contingency B: (a) unstable under original OPF; (b) stable under RTSC-OPF-PFR}
 \label{fig:L23-24 stable}
\end{figure}

\begin{figure}[t]   
 \includegraphics[width=\linewidth]{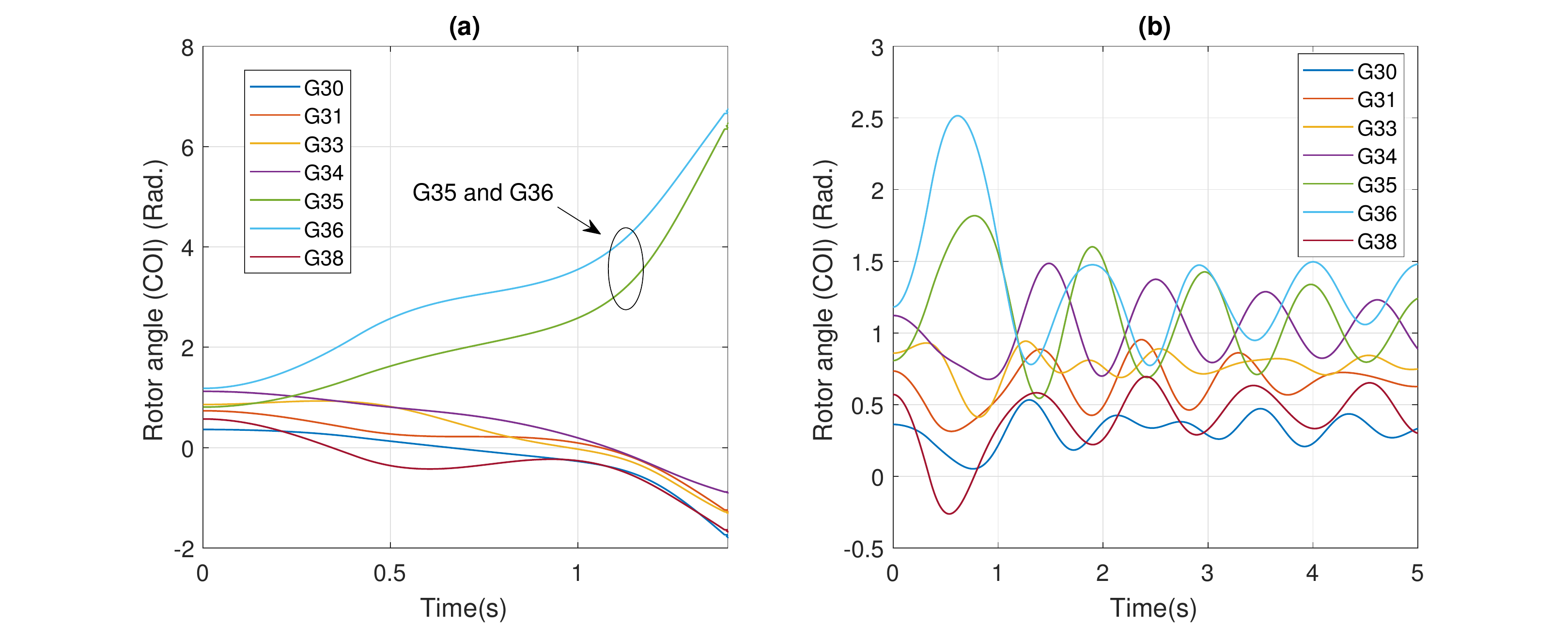} 
 \caption{Rotor angle trajectories under Contingency C: (a) unstable under original OPF; (b) stable under RTSC-OPF-PFR}
 \label{fig:L21-22 stable}
\end{figure}

We test the robustness of the proposed scheme to three contingencies for the modified New England 39-Bus system. The corresponding transient stability constraints are included in the optimization problem so that the obtained solution is stable under each of the contingencies.  
The details of the contingencies are as follows:

\begin{itemize}
\item Contingency 1-A three-phase to ground fault is considered at bus 29 and cleared by tripping Line 26--29 at 0.350~s, which is greater than the initial CCT 0.310 s.

\item Contingency 2-A three-phase to ground fault is considered at bus 23 and cleared by tripping Line 23--24 at 0.350~s, which is greater than the initial CCT 0.313 s.

\item Contingency 3-A three-phase to ground fault is considered at bus 22 and cleared by tripping  Line 21--22 at 0.470~s, which is greater than the initial CCT 0.457 s.

\end{itemize}


To have a fair comparison, five approaches are tested, including the normal OPF, TSC-OPF, TSC-OPF-PFR, robust TSC-OPF without PFRs, and the proposed RTSC-OPF-PFR. For the approaches without considering the renewable uncertainties, the mean values of the wind power outputs are used. 
The results of the different approaches are depicted in Table \ref{table:2}. Although the normal OPF gives a solution with the lowest generation cost, its robustness under different contingencies shows that the system becomes unstable in most of the wind power scenarios.  For TSC-OPF and TSC-OPF-PFR, the robustness under the C1-C3 increases when compared to the normal OPF. This means that the transient stability constraints force the system to adjust the power transfer between the critical and non-critical generators and increase the declaration area under the SIME-based analysis. TSC-OPF-PFR gives a better performance on robustness with lower generation cost, which indicates the value of PFRs in enlarging the feasible region of OPF problems. However, transient instability still occurs in a considerable portion of wind power scenarios. This infers that these two approaches without considering the wind power variations are still vulnerable to the contingencies.

To avoid the high possibility of transient instability under uncertainties, two robust approaches, i.e., RTSC-OPF and RTSC-OPF-PFR, are tested. We can see from Table \ref{table:2} that the performance of robustness and CCTs under the two robust approaches are much better than the previous three methods. 
For system robustness, RTSC-OPF-PFR is more robust than RTSC-OPF and the generation cost of RTSC-OPF-PFR is 3.0\% (851.28\$/hr) lower than RTSC-OPF.
For CCTs, it should be noted that although the CCTs of TSC-OPF and TSC-OPF-PFR are at least 200 ms more than that of the normal OPF, the low robustness of these two methods indicates that its validity become doubtful if variable wind power outputs are considered. This is quite different from the proposed approach where the high robustness level suggests the high reliability of the CCTs.  
It is worth mentioning that RTSC-OPF-PFR achieves very high robustness level (almost 100\%) with the generation cost being only 1.15\% higher than TSC-OPF. It means that the system can obtain stability robustness in a more economy way with the inclusion of PFRs.

In Figs. \ref{fig:L26-29 stable}-\ref{fig:L21-22 stable}, unstable and stable cases under C1-C3 are shown with multi-machine rotor angles under the original OPF and the proposed RTSC-OPF-PFR, respectively.  

Under the proposed framework, the obtained solution achieves high economy and high robustness regarding transient stability. The computation time is within one minute so it can be applied in online dispatch. Moreover, the proposed optimization model exploits the value of PFR in reducing generation cost and enhancing system stability under renewable uncertainties. PFR is also expected to provide more flexibility for system to accommodate higher penetration of RESs  while bringing good economic merit.

\section{Conclusion} \label{sec:conclusion}
This paper proposes RTSC-OPF-PFR problem to handle system operation under renewable uncertainties. There are two main foci regarding the optimization model and solution framework.
Firstly, we introduce a network controller, called the PFR, into the optimization problem which provides a network-side capability for cost reduction while ensuring system stability. Then we design an offline-online solution framework which enables the RTSC-OPF-PFR for real-time implementation. Under the methods and assumptions in Section \ref{sec:methods}, the original RO problem can be simplified into a low-dimensional deterministic problem, which can be solved efficiently for online dispatch. Numerical tests on the modified New England 39-bus system show that the proposed method coordinating PFRs and generation re-dispatch can significantly improve the robustness level at the expense of low generation cost which is superior to the traditional TSC-OPF methods that consider the generation re-dispatch only. Also, the computation time of the problem under the proposed solution framework is quite low which is suitable for practical real-time dispatch.

There are many directions which can be considered for future research. For example, chance constraints can be our future work to model the uncertainties of RESs. Load uncertainties and more generic load models need to be further developed. Moreover, corrective control could be considered combined with the preventive control to improve the system transient stability.

\section*{Acknowledgment} 
This research is supported by the Theme-based Research Scheme of the Research Grants Council of Hong Kong, under Grant No. T23-701/14-N.

\ifCLASSOPTIONcaptionsoff
  \newpage
\fi

\bibliographystyle{IEEEtran}

\bibliography{Reference,IEEEabrv}

\end{document}